# Health-related Quality of life, Financial Toxicity, Productivity Loss and Catastrophic Health Expenditures After Lung Cancer Diagnosis in Argentina.

Lucas Gonzalez [1,2]; Andrea Alcaraz[1]*; Carolina Gabay[3]; Mónica Castro[3]; Silvina Vigo[4]; Eduardo Carinci[4]; Federico Augustovski[1,5].

1. Institute for Clinical Effectiveness and Health Policy (IECS), Dr. Emilio Ravignani 2024 Buenos Aires, Argentina.
2. Hospital Interzonal General de Agudos Profesor Dr. Rodolfo Rossi, Oncology Unit, Calle 37 No 183 La Plata, Argentina.
3. Instituto de Oncología Ángel H. Roffo, Thoracic Oncology Functional Unit, Av. San Martín 5481, Buenos Aires, Argentina.
4. Hospital Interzonal Especializado de Agudos y Crónicos San Juan de Dios, Oncology Unit, Calle 27 y 70 La Plata, Argentina.
5. Centro de Investigaciones en Epidemiología y Salud Pública (CIESP), Dr. Emilio Ravignani 2024, Buenos Aires, Argentina.

\*   Corresponding author

**ABSTRACT**

*Objective:*
About 12,000 people are diagnosed with lung cancer (LC) each year in Argentina, and the diagnosis has a significant personal and family. The objective of this study was to characterize the Health-Related Quality of Life (HRQoL) and the economic impact in patients with LC and in their households.

*Design:*
Observational cross-sectional study, through validated structured questionnaires (self-administered interview).

*Setting*
Two referral public health care centers in Argentina.

*Participants*
Adult patients with a diagnosis of Non-Small Cell Lung Cancer (NSCLC).

*Main outcomes*
Health-Related Quality of life (EuroQol EQ-5D-3L questionnaire); financial toxicity (COST questionnaire); productivity loss (WPAI-GH, Work Productivity and Activity Impairment Questionnaire: General Health), and out-of-pocket expenses.

*Results:*
We included 101 consecutive patients (mean age 67.5 years; 55.4% men; 57.4% with advanced disease -stage III/IV-). The mean EQ-VAS was 68.8 (SD:18.3), with 82.2% describing fair or poor health. The most affected dimensions were anxiety/depression, pain, and activities of daily living. Patients reported an average 59% decrease in their ability to perform regular daily activities as measured by WPAI-GH. 54.5% reported a reduction in income due to the disease, and 19.8% lost their jobs. The annual economic productivity loss was estimated at USD 2,465 per person. 70.3%) of patients reported financial toxicity. The average out-of-pocket expenditure was USD 100.38 per person per month, which represented 18.5% of household income. Catastrophic expenditures were present in 37.1% of households. When performing subgroup analysis by disease severity, all outcomes were worse in the subpopulation with advanced disease.

*Conclusions*
Patients with NSCLC treated in public hospitals in Argentina have significant health-related quality of life and economic impact, worsening in patients with advanced disease.

**Strengths and limitations of this study**

- This study offers a unique opportunity to study health and economics outcomes on an individual level in patients affected by lung cancer, including both the individual and family's economic and social perspective.
- We utilized standardized and validated instruments to measure health-related quality of life, financial toxicity, and work productivity, enabling comparability with other studies.
- It is to our knowledge the first study that has this scope in Argentina and Latin America.
- The main limitations pertain to the fact that the study was not conducted on a representative sample of patients from Argentina, and was done in public institutions, thus precluding conclusions at a national level.

**Funding**

'This work was supported by National Cancer Institute, grant number DI-2018-19-APN-INC # MS.

**Competing interests.**

The authors declare that they have no competing interests.

**Word count for the manuscript:** 2928.

# 1. INTRODUCTION

Lung cancer (LC) is the most frequently diagnosed tumor worldwide. In 2020, it constituted 12.2% (2.2 million) of all diagnosed tumors and caused 18. 1% (1.7 million) of deaths, ranking as the leading cause of death from cancer[1, 2].

Argentina showed an incidence of 12,110 annual cases in 2020, data that place it in fourth place for both sexes. It is the leading cause of cancer mortality in our country, representing 10,729 deaths in 2020. [3]

Non-small-cell lung carcinoma (NSCLC) accounts for about 85% of all lung cancers.[4] Patient survival is closely related to the stage of the disease at the time of diagnosis. Unfortunately, most patients (85%) have an advanced detection stage. [5, 6].

The diagnosis of NSCLC causes a profound impact on the family and the individual's social network. It involves modifying the natural course of personal life and family dynamics. The notice of the disease, the change in the perception of life expectancy, and the presence of symptoms associated with the disease or the treatments performed, among others, cause an undeniable deterioration in the health-related Quality of life (HRQoL).[7-9].

Likewise, many households are economically affected by the direct costs generated during the diagnostic and therapeutic process or by the loss of productivity. Although medical coverage can cushion the economic impact concerning some expenses, fundamentally the most expensive ones related to treatment (such as medication or studies), other expenses such as transportation, caregivers, and productivity costs (absenteeism), are not usually contemplated. [10]

This situation naturally worsens among those without coverage or low income because they not only have to face a situation of access to benefits and appropriate resources but also their family members can lose hours of work as medical care (unpaid) or even increase them to obtain higher incomes to face the high expenses associated with the disease. The productivity of household members is affected by the presence of a family member with the disease. [11] Thus, this population is more vulnerable since the situation they face can be catastrophic for the economy of their homes.

"Financial toxicity" is a term that describes the problems faced by cancer patients for the costs associated with the treatment you must pay self-employment (pocket expenses or "out-of-pocket"). Multiple studies indicate that patients with a diagnosis of cancer who suffer from financial toxicity are more likely to have less adherence to treatment or follow-up, which affects the effectiveness of treatments with a detriment to their survival and Quality of life.).[12-14]

There are no local studies published in peer-reviewed journals that portray the economic impact of NSCLC on patients and their homes and compare it in different population groups such as high-, middle- or low-income groups or people with and without formal health coverage (public, private or social security plan).

This study aimed to characterize health-related Quality of life and economic impact in patients with a recent diagnosis of lung cancer who perform their treatment in public hospitals in Argentina through different validated instruments to measure HRQoL, loss of labor productivity, and financial stress. The hypothesis was that lung cancer could have a negative impact on these outcomes.

# 2. METHODS

A cross-sectional study was conducted with collection of clinical, demographic, socioeconomic, and out-of-pocket data at the individual and household levels in two public hospitals in Argentina (Instituto de Oncología Ángel H. Roffo and Hospital Interzonal Especializado de Agudos y Crónicos San Juan de Dios). The institutional research ethics committee of Instituto de Oncología Ángel H. Roffo approved the study. Informed consent was obtained from all participants.

Selection criteria

All patients older than 18 years who attended a scheduled appointment were eligible for participation. Inclusion criteria were: older than 18 years, pathologically confirmed diagnosis of NSCLC and more than three months from diagnosis. Those who were participating in a therapeutic intervention study, unable to provide informed consent, or were unable to read, understand, and complete were excluded.

Clinical, demographic, socioeconomic, and out-of-pocket data at the individual and household levels were collected through a specifically designed questionnaire completed in the form of an interview. All the instruments were administered after all patients' informed consent was obtained and signed.

*Health-Related Quality of Life*

HRQoL was assessed using the EQ-5D3L instrument adapted and validated for Argentina. [15] The instrument consists of five items assessing five dimensions: mobility, self-care, activities of daily living, pain/discomfort, and anxiety/depression. Each question has three response options ranging from 1 'I have no problems' to 3 'I have many problems.' Additionally, patients are asked to describe their health status on a Visual Analogue Scale (EQ-VAS), ranging from 0 (worst imaginable health status) to 100 (best imaginable health status). For each health state described, the EQ-5D Index of preference values was obtained from a social valuation study in the general population in Argentina using the time trade-off as a valuation technique.[15, 16]

*Productivity*

The effects of NSCLC diagnosis and treatment on daily activities and productivity were assessed using the validated WPAI (Work Productivity And Activity Impairment) instrument in its General Health (GH) version.[17]

The WPAI questionnaire also allows the effect of the disease on productivity to be transformed into a monetary value by multiplying the values obtained in the test by the hourly wage of the patient interviewed, thus estimating the economic loss caused by the disease. To estimate the economic loss, the average monthly income declared by the patient was annualized and then divided by the number of annual working hours based on a 40-hour working week and 12 annual holidays. Personal income was not assessed precisely but by categories, so for the calculation; it was assumed that the income of each participant was the average value of each category.

*Out-of-pocket expenditures*

Out-of-pocket (OOP) expenditures at the individual and household level were estimated globally and through a micro-costing approach. Patients were asked to consider health expenditures in the last month through the first method. Then we asked to include those expenditures incurred on transfers, meals away from home, medicines, other treatments, tests, and caregivers, among others. Using the micro-costing approach, they were estimated according to different health resources used in the last month, their amounts, utilization rates, and individual value declared for each of them. The categories of health resources included were the number of scheduled consultations, number of unscheduled consultations (emergency or on-call), the performance of studies or analyses, the performance of administrative procedures, purchase of medicines, expenses incurred for surgeries, radiant treatment, or others. The reimbursements made by the pre-payers were estimated and deducted from the total expenditure. In this way, it was possible to establish the structure of out-of-pocket expenditure.

*Financial toxicity*

The patient's emotional state concerning the financial situation was assessed using the validated COST instrument in its Spanish version (Comprehensive Score for Financial Toxicity).[19-21] The instrument consists of 11 items/questions, each with multiple response options ranging from 0 to 4. The resulting score can range from 0 to 44 (the lower the score, the greater the financial toxicity or impact on HRQoL), with values above 26 considered "normal" or with no impact on Quality of life. Values below this threshold can be subdivided into degrees of severity, whereby a score between 14 and 25 means mild impact (Grade 1), 1 to 13 means moderate impact (Grade 2), and 0 (zero) means high impact (Grade 3). According to these criteria, the presence of grade 1 or higher (COST < 26) was defined as financial toxicity.

*Catastrophic expenditure*

To assess the prevalence of catastrophic health expenditure related to the diagnosis of NSCLC, we used two definitions: a) if the reported out-of-pocket health expenditure in the last month for care was 10% or more of the total reported household income; or b)if the reported out-of-pocket health expenditure in the last month for the care of the person diagnosed with NSCLC was 40% or more of the household's total ability to pay. The ability to pay was defined as the actual household income not spent on housing and food. According to the National Household Expenditure Survey 2017-2018, it was estimated that 37.2% of household income is spent on housing and food, so the ability to pay was the result of subtracting this percentage from the average reported household income.

*Statistical Analyses*

The patient data was anonymously loaded into the RedCap computer system, which was preloaded with a validation system to minimize missing data. In the event of missing data, the original forms were reviewed again. If the data was still missing, it was reported as a missing data point. The patient data were Descriptive statistics summarize the baseline demographic, HRQoL, work productivity, and household financial status variables. Between-group comparisons of data were analyzed using Student's t-test, one-way ANOVA, or Kruskal Wallis H-test according to the distribution of variables for continuous data, and chi-square or Fisher's exact for categorical data, as appropriate. Correlations were assessed using the Pearson correlation. The correlation was considered mild if the coefficient was between 0.20 and 0.39, moderate if between 0.40 and 0.59, strong if between 0.60 and 0.79, and very strong if 0.80. The STATA® 16 toolkit was used for data analysis. Results with an alpha level of less than 0.05 were interpreted as significant.

3. **RESULTS**

Out of 249 scheduled and consecutive consultations, 101 patients met the inclusion criteria, signed the informed consent, and completed the questionnaires between May 2nd and December 16th, 2019. Two patients were excluded as they did not complete the questionnaire before completion of recruitment and had given their consent.

The baseline patient socio-demographic and the tumor characteristics are summarized in Table 1. The mean time between diagnosis and interview was 920 days (95% CI 720 - 1121).

| Table 1 |
|---|

*3.1 Health-Related Quality of Life in patients with lung cancer.*

The mean EQ-VAS was 68,8 (SD:18,3). There was no difference in HRQoL when comparing sex, stage of disease or when clustering patients into early (stages I, II, IIIa) and advanced (stages IIIb and IV) disease (see Table 2). No differences were found when using the EQ-5D Index. Patients described 34 different health states.

| Table 2 |
|---|

Figure 1 shows the degree of impairment in each of the domains of the EQ-5D-3L questionnaire. The disease affected all the dimensions, with a significant impact on daily activities and mobility. More than half of the patients reported being anxious or depressed and uncontrolled pain.

| Figure 1 |
|---|

*3.2 Productivity in patients with lung cancer*

According to the WPAI instrument, patients diagnosed with NSCLC reported an average 59% decrease in their ability to perform regular daily activities.

Nearly 17% (17/101) of the population was active. In this population, the total average number of working hours lost per week was 15.35 hours (SD ± 15.03), a percentage of working time lost due to absenteeism of 61.12% (SD

± 32.9%). While the decrease in performance during the hours worked (presenteeism) was 54.7%. Therefore, the total productivity loss reached 72.2% (Table 3 and Supplementary Information: Table A).

|         |
|---------|
| Table 3 |

The monetary loss due to absenteeism could only be estimated for 14 persons who additionally reported personal income values. It was estimated at USD 2,465 (95% CI: 780 -4,150) person-years due to absenteeism raised to USD 2,622 (95% CI: 966 – 4,277) considering presenteeism. Statistically, significant differences were observed between patients with advanced and early disease due to absenteeism only (USD 648 vs. 3827; p:0.01) or absenteeism and presenteeism (USD 735 vs. 4037; p<0.01).

This study also asked whether they are currently working and other questions to assess family and household impact beyond reducing working hours. Table 3 shows other measures of work impact in NSCLC patients.

*3.3 Out-of-pocket expenditures*

The mean total OOP expenditures reported by participants was USD 100.38. Table 4 shows the average monthly expenditure reported for the entire population and disease stage.

|         |
|---------|
| Table 4 |

When assessing the structure of out-of-pocket expenditure, the primary inputs were pharmacy and scheduled or urgent doctor visits (see Supplementary Information: Table B).

The percentage of health expenditure was 18.5% of household income. Catastrophic health expenditure could be estimated in 66.3% of the sample (65/101) who reported out-of-pocket expenses and family income. Table 4 shows the estimated prevalence of catastrophic expenditure in our study population.

*3.4 Financial toxicity.*

The median COST score was 20 (range 2 to 40; mean, 20.1 SD = 9.94; 95% CI: 18.2 - 22.1). About 70.3% of the patients reported some degree of financial toxicity. Patients with advanced disease were more likely to present financial toxicity than those with early stages (Table 2 and Supplementary Information: Table C and D).

HRQoL had a moderate and statistically significant correlation with the financial toxicity reported by the patients (0.47 with the EQ-VAS and 0.39 with the EuroQol Index).

4. **DISCUSSION**

In Argentina, lung cancer is the leading cause of cancer-related deaths. This is the first study that assesses lung cancer's impact on health-related Quality of life, financial toxicity, work productivity, and catastrophic health expenditure in a large sample of patients served by the public sector of the city and province of Buenos Aires. Our main results show that this patient population with low socioeconomic status has poor health-related Quality of life and a substantial decline in their ability to perform regular daily activities as measured by WPAI-GH (of nearly 60%), with a consequent decline in income and job loss. More than two-thirds of the sample reported some degree of financial toxicity, and disease-related costs represented roughly 20% of household income. The average annual monetary loss calculated was USD 2622 per person per year, higher for inactive patients with advanced stages of the disease (AR$ 191,175 versus AR$ 34,808; p<0.01). Catastrophic expenditures related to the disease were observed in 24% to 54% of the sample when using a more stringent or sensitive threshold of more than 40% or 10% of household income. One in five patients had to borrow money, and 2 in 5 had to suspend planned projects. The impact on health-related Quality of life, financial toxicity, work productivity, and catastrophic health expenditure was more significant in patients with advanced disease, though to the smaller subgroup samples was not always statistically significant.

Comparison with the observed HRQoL of the general Argentine population shows an evident deterioration in patients with non-small cell lung cancer: while the general argentine population has a high HRQol, with a 0,902 mean EQ-Index; the mean HRQol in our sample was significantly and clinically lower (mean EQ-Index 0.72), and similar to those who report fair or poor health.[16] Although we found no evidence from Argentina, our findings are in line with recent evidence that shows not only the negative impact of NSCLC in HRQoL but also in all the financial and economic domains (i.e., catastrophic spending, financial burden, and toxicity), and that these effects are generally more pronounced in more advanced disease. Different systematic reviews in NSCLC show the negative impact on HRQoL: one study that included 19 studies showed a negative impact in both mental and physical domains in surgical patients; another one focused on randomized trials (N=53) and that even though 8 in 10 did not show a significant effect on overall survival; in five out of ten this "negative" studies, a significant effect in HRQOL was observed.[8,21] Financial toxicity is a relatively novel concept that helps guide policy actions.[22] It has shown to be related to unemployment in patients with cancer, increased overall and especially psychologic symptom burden, and even shorter patient survival.[23-25] A recent systematic review that assessed this issue among patients with cancer in publicly funded healthcare countries and included 32 articles showed that even in government-funded universal public healthcare systems, financial toxicity is an important issue that may need financial protection measures. This financial impact can be even more critical in countries such as Argentina without universal publicly funded healthcare systems.

Our study has several strengths as well as limitations. It is the first study performed in Argentina to assess health-related Quality of life, financial toxicity, work productivity, and catastrophic health expenditure in patients served by the public sector of the city and province of Buenos Aires, using a rigorous protocol and cross-sectional methodology, that helps to depicts these relevant aspects in this patient population. Also, our protocol allowed us to avoid recruitment bias by including a consecutive sample.

The main limitations relate to the fact that the study was not performed in a representative sample of patients from Argentina and thus cannot draw conclusions on Argentina as a whole: a) the study sample was composed of a population served by the public healthcare sector of two referral centers of the largest country's jurisdictions (city and province of Buenos Aires, that compose around 45% of the country's population); and b) the study was performed in two public hospitals (and the public sector serves around 40% of the whole country's population (with 50% served by the social security system and 10% by the private healthcare sector).[26] An extension of the present study to the private sector is currently being carried out. It will help understand the impact of non-small cell lung cancer in this specific population and compare it to those served by public institutions.

In summary, our study highlights the relevance and impact of non-small cell lung cancer on health-related Quality of life, financial toxicity, work productivity, and catastrophic health expenditure in a large sample of patients served by the public sector of the city and province of Buenos Aires. It emphasizes the need to recognize and develop tools to improve well-being and reduce the financial burden by the care teams and healthcare system. Further studies in other healthcare sectors and other jurisdictions will contribute to more broadly assessing the impact of non-small cell lung cancer in Argentina.

**List of Abbreviations.**
LC: Lung Cancer; HRQOL: Health-related quality of life; OOP: out-of-pocket; NSCLC: Non-small-cell lung carcinoma VAS: Visual Analogue Scale WPAI: Work Productivity And Activity Impairment SD: Standard deviation COST: Comprehensive Score for Financial Toxicity USD: US dollar.

**Declarations Ethics approval and consent to participate.**
The Research and Ethics Committe of the Instituto de Oncología Ángel H. Roffo granted ethical approval for this study and the study adhered to the tenets of the Declaration of Helsinki. Informed consent was obtained from all participants.

**Consent for publication.**

Not applicable.

**Patient and Public Involvement statement**

Patients were initially involved in the research as participants in the study. The outcome measures were developed using standardized questionnaires that were guided by their priorities, experience, and preferences. Patients were not involved in the design, recruitment or conduct of this study, which nevertheless adhered to the research guidelines of the National Cancer Institute in Argentina. They were informed in order to assess the burden of the time required to participate in the research when they signed the consent form, approved by the Internal Review Board that approved the study Patients were not involved in the plans to disseminate the study results to participants and relevant patient communities.

**Availability of data and materials.**

The datasets used and/or analyzed during the present study are not publicly available due to the General Data Protection Regulation laws but are available from the corresponding author on reasonable request and with permission from.

**Author contributions.**

L.G, A.A and F.A. developed the conception and wrote the protocol.

C.G, M.C, S.V, and E.C, recruited and interviewed patients.

L.G and A.A collected, analyzed, and interpreted data.

L.G, A.A and F.A. wrote the main manuscript text and tables.

All authors reviewed and approved the final manuscript.

**Acknowledgements.**
Not applicable.

**Table 1. Characteristics of lung cancer patients**

| Socio-demographic characteristics | Central Estimate | Deviation |
|---|---|---|
| Mean Age (years) | 65,7 | SD 8,9 |
| Mean time to diagnosis (days) | 920 | SD 1009 |
| Median time to diagnosis (days) | 609 | IQR 276 - 1069 |
| **Sex** | **N** | **Value** |
| Male | 56 | 55,4% |
| Female | 45 | 44,6% |
| **Medical Center** | | |
| Instituto de Oncología Ángel H. Roffo | 76 | 75,2% |
| H.I.E.A y C. San Juan de Dios | 25 | 24,8% |
| **Residence** | | |
| CABA | 17 | 16,8% |
| Buenos Aires Province | 74 | 73,3% |
| Otro | 10 | 9,9% |
| **Marital status** | | |
| Single | 9 | 8,9% |
| Married | 50 | 49,5% |
| Concubine | 13 | 12,9% |
| Widowed | 14 | 13,9% |
| Divorced/Separated | 15 | 14,8% |
| **Education** | | |
| Incomplete primary school | 14 | 13,9% |
| Primary school completed | 37 | 36,6% |
| Incomplete secondary education | 15 | 14,8% |
| Secondary school completed | 19 | 18,8% |
| Tertiary/University | 16 | 15,8% |
| **Medical insurance *** | | |
| PAMI (retirees) | 56 | 55,4% |
| Employer insurance (social security) | 27 | 26,7% |
| Private | 3 | 3,0% |
| Public | 19 | 18,8% |
| **Household income (USD)** | | |
| ≤ 80 | 5 | 4,9% |
| >80 and ≤ 122 | 2 | 1,9% |
| >122 and ≤ 338 | 8 | 7,9% |
| >338 and ≤ 412 | 6 | 5,9% |
| >412 and ≤ 486 | 13 | 12,9% |
| >486 and ≤ 591 | 8 | 7,9% |
| >591 and ≤ 718 | 11 | 10,9% |
| >718 and ≤ 940 | 16 | 15,8% |
| >940 and ≤ 1191 | 5 | 4,9% |
| >1191 | 8 | 7,9% |
| Not reported | 19 | 19,1% |
| Stage of disease | | |
| I | 13 | 12,9% |
| II | 15 | 14,8% |
| IIIa | 15 | 14,8% |
| IIIb | 16 | 15,8% |
| IV without brain metastasis | 42 | 32,2% |
| IV with brain metastasis | 4 | 9,5% |
| ECOG performance status | | |
| 0-I | 34 | 3,7% |
| II | 55 | 54,.5% |
| III | 12 | 11,9% |
| IV | 0 | 0,0% |

*Note: \* Some patients report more than one medical insurance*

*PAMI: Comprehensive Medical Care Program is an insurance for retirees and pensioners*

*Mean exchange rate USD 1 = ARS 47,36*

**Table 2. Health related quality of life, financial toxicity and disability in patients with lung cancer in Argentina.**

| Indicator | Global N=101 Mean | CI | Early Stage (I-II-IIIa) N=43 Mean | CI | Advanced Stage (IIIb-IV) N=58 Mean | CI |
|---|---|---|---|---|---|---|
| Quality of life: EQ VAS-5D-3L score | **68,8** | 65,2 - 72,4 | **70,9** | 65,6 - 76,2 | **67,8** | 62,9 - 72,7 |
| Quality of life: EQ-5D Index | **0,72** | 0,68 - 0,76 | **0,76** | 0,69 - 0,83 | **0,7** | 0,65 - 0,75 |
| COST questionnarie value | **20,1** | 18,2 - 22,1 | **24** | 21,3 - 26,7 | **17,3** | 14,7 - 19,9 |
| Prevalence of Financial Toxicity by COST questionnarie | **70,3%** | 60,4 - 78,98 | **51,2%** | 35,5 - 66,7 | **84,5%** | 72,6 - 92,7 |
| Percentage of disability or impairment of daily activities due to the disease By WPAI questionnarie | **59,0%** | 53,2 – 64,8 | **53,7%** | 44,6 – 62,8 | **62,9%** | 55,3 – 70,5 |

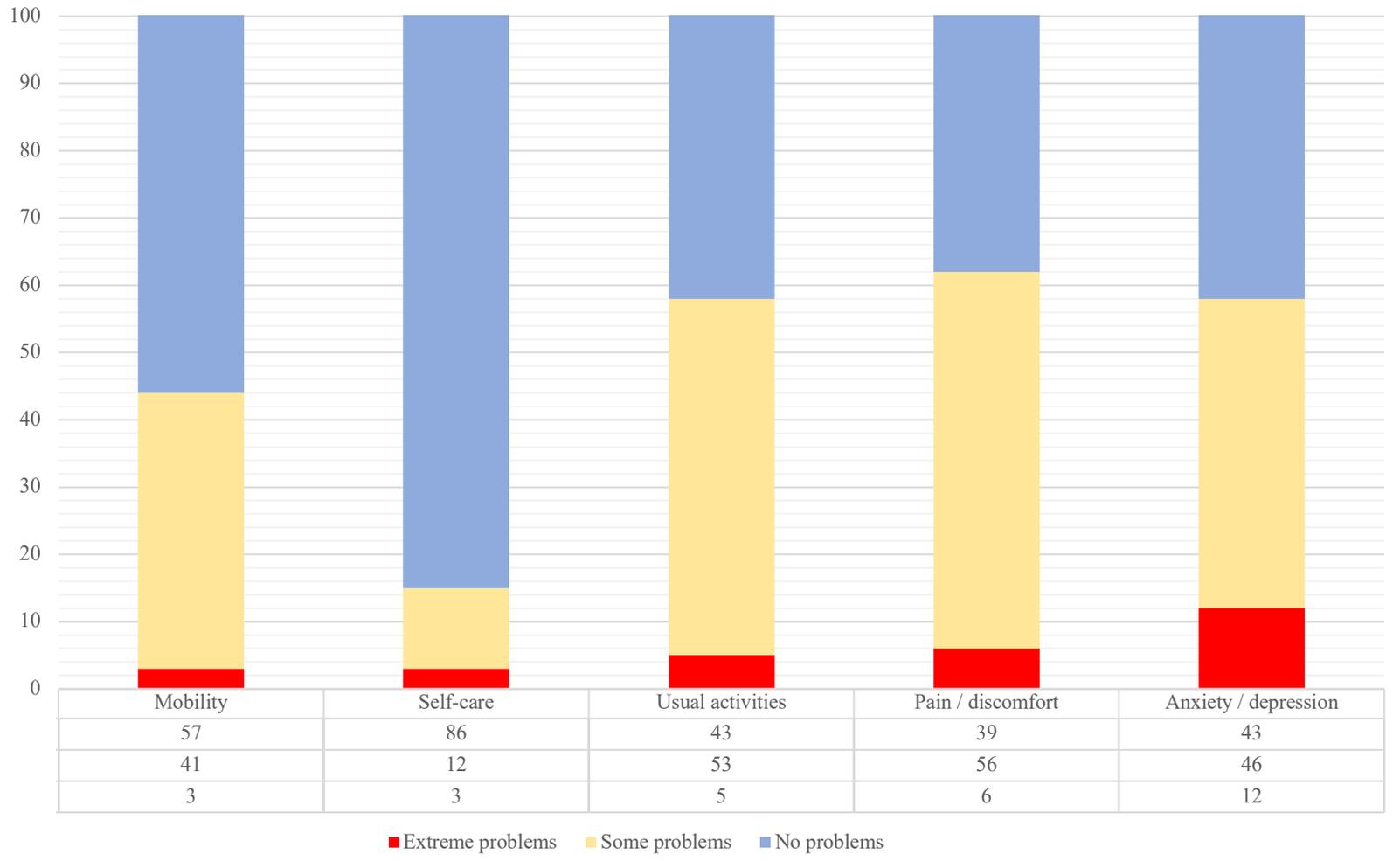

**Table 3. Patient productivity impact and social family impact of lung cancer.**

| Laboral impact. General questions | 101 | Value |
|---|---|---|
| Are you currently working? | 25 | 24,7% |
| Was your income from your work (salary, salary, money) reduced because of your illness? | 55 | 54,5% |
| Did you have to resign or get fired because of a lung cancer diagnosis? | 20 | 19,8% |
| Did you have to change jobs because of your illness? | 16 | 15,8% |
| Laboral impact. WPAI questionnaire. * | 17 | |
| Hours per week lost due to illness | 17 | 15,35 |
| Hours per week lost due to other causes | 17 | 0 ,94 |
| Absenteeism | 17 | 61,12 |
| Presentism | 17 | 54,7 |
| Total productivity loss | 17 | 76,2 |
| Annual monetary loss due to absenteeism | 17 | USD 2.465 |
| Annual monetary loss due to absenteeism and presenteeism | 17 | USD 2.622 |
| Family and household impact | 101 | Value |
| Has any family project had to be suspended or canceled for financial reasons because of your illness? | 39 | 38,6% |
| Have you had to sell (you and your family) things, for example a car, land, goods, etc., because of your illness? | 12 | 11,9% |
| Are there any medical indications that you could not comply with for financial reasons? | 7 | 6,9% |
| Did your family have to borrow or are you thinking of borrowing money because of lung cancer? | 22 | 21,8% |
| Adding up all the earned income for you and the family that lives with you, was it reduced because of cancer? | 56 | 57,1% |

*Note: * Only in working population*

*Mean exchange rate USD 1 = ARS 47,36*

**Table 4. Economic impact of lung cancer.**

| | N | Central value | IC 95% | |
|---|---|---|---|---|
| | | | Low | High |
| *Monthly out-of-pocket spending on health (USD)* | | | | |
| Global | 80 | USD 100,38 | USD 60,49 | USD 140,27 |
| Early Stage (I-II-IIIa) | 35 | USD 50,09 | USD 25,09 | USD 75,09 |
| Advanced Stage (IIIb-IV) | 45 | USD 139,48 | USD 72,30 | USD 206,67 |
| *Percentage of family income earmarked for health.* | | | | |
| Global | 67 | 18,5% | 12,6% | 24,5% |
| Early Stage (I-II-IIIa) | 29 | 12,6% | 5,8% | 19,4% |
| Advanced Stage (IIIb-IV) | 38 | 23,1% | 13,9% | 32,2% |
| *Prevalence of catastrophic spending: Health spending> 10% of family income* | | | | |
| Global | 26/70 | 37,1% | 25,9% | 49,5% |
| Early Stage (I-II-IIIa) | 9/31 | 29,0% | 14,2% | 48,0% |
| Advanced Stage (IIIb-IV) | 17/39 | 43,6% | 27,8% | 60,4% |
| *Prevalence of catastrophic spending: Health spending> 40% of ability to pay* | | | | |
| Global | 16/70 | 22,9% | 13,7% | 34,4% |
| Early Stage (I-II-IIIa) | 5/31 | 16,1% | 5,5% | 33,7% |
| Advanced Stage (IIIb-IV) | 11/39 | 28,2% | 15,0% | 44,9% |